\documentclass[nocompress]{spie}

\usepackage{amsmath,amsfonts,amssymb}
\usepackage{graphicx}
\usepackage[colorlinks=true, allcolors=blue]{hyperref}

\usepackage{subcaption}
\usepackage{multirow}
\usepackage{bm}
\usepackage[labelfont=bf, skip=8pt]{caption}
\usepackage[indLines=false]{algpseudocodex}

\usepackage{fancyhdr}
\usepackage{booktabs}

\newcommand{\R}{\mathbb{R}}
\newcommand{\E}{\mathbb{E}}
\newcommand{\x}{\bm{x}}

\def\PAnumber{AFRL-2025-4305}

\fancypagestyle{empty}{%
  \fancyhf{} 
  \fancyfoot[C]{Approved for public release; distribution is unlimited.  Public Affairs release approval \# \PAnumber .}
  
}

\title{Boiling flow parameter estimation from boundary layer data}

\author[a]{Jeffrey W. Utley}
\author[a]{Gregery T. Buzzard}
\author[b]{Charles A. Bouman}
\author[c]{Matthew R. Kemnetz}
\affil[a]{Department of Mathematics, Purdue University, West Lafayette, Indiana 47907, USA}
\affil[b]{Departments of Electrical and Computer Engineering, and Biomedical Engineering, Purdue University,  West Lafayette, Indiana 47907, USA}
\affil[c]{Department of Engineering Physics, Air Force Institute of Technology, Wright-Patterson AFB, OH 45433, USA}

\authorinfo{Further author information: (Send correspondence to J.W.U.)\\J.W.U.: E-mail: utleyj@purdue.edu}

\begin{document} 
\maketitle

\begin{abstract}
Atmospheric turbulence and aero-optic effects cause phase aberrations in propagating light waves, thereby reducing effectiveness in transmitting and receiving coherent light from an aircraft. Existing optical sensors can measure the resulting phase aberrations, but the physical experiments required to induce these aberrations are expensive and time-intensive. Simulation methods could provide a less expensive alternative. For example, an existing simulation algorithm called boiling flow, which generalizes the Taylor frozen-flow method, can generate synthetic phase aberration data (i.e., phase screens) induced by atmospheric turbulence. However, boiling flow depends on physical parameters, such as the Fried coherence length $r_0$, which are not well-defined for aero-optic effects. In this paper, we introduce a method to estimate the parameters of boiling flow from measured aero-optic phase aberration data. Our algorithm estimates these parameters to fit the spatial and temporal statistics of the measured data. This method is computationally efficient and our experiments show that the temporal power spectral density of the slopes of the synthetic phase screens reasonably matches that of the measured phase aberrations from two turbulent boundary layer data sets, with errors between 8-9\%. However, the Kolmogorov spatial structure function of the phase screens does not match that of the measured phase aberrations, with errors above 28\%. This suggests that, while the parameters of boiling flow can reasonably fit the temporal statistics of highly convective data, they cannot fit the complex spatial statistics of aero-optic phase aberrations.
\end{abstract}


\section{INTRODUCTION}
\label{s: Introduction}

Atmospheric turbulence and aero-optic effects distort light wave propagation, thereby reducing effectiveness in transmitting and receiving coherent light from an aircraft. In particular, both phenomena cause refractive index variations \cite{Tatarski, WangPhysicsComputation, Fitzgerald} which lead to random phase aberrations \cite{Visbal, Chernov, Kalensky}. These spatially and temporally varying phase aberrations can be measured by optical sensors \cite{Kemnetz, Geary, Holmes}, but the physical experiments required to measure this data are expensive and time-intensive \cite{JumperAAOL}. Simulation methods could provide a less expensive alternative. Existing simulation algorithms using the Kolmogorov theory of turbulence and the Taylor frozen-flow hypothesis can generate synthetic two-dimensional phase aberration data, called phase screens, induced by atmospheric turbulence \cite{Schmidt, Srinath, Vogel}; phase screens generated using the Kolmogorov theory of turbulence are often called random Kolmogorov phase screens. However, these simulation algorithms depend on physical parameters such as the wind velocity components $(v_x,v_y)$, outer scale $L_0$, and Fried coherence length $r_0$. Because aero-optic effects do not follow the Kolmogorov theory of turbulence \cite{Vogel, Siegenthaler}, the physical parameters $L_0$ and $r_0$ may be undefined for aero-optic effects.

Boiling flow is a prominent phase screen generation method which generates time-series of random Kolmogorov phase screens. Introduced by Srinath \textit{et al.} \cite{Srinath}, this method generalizes the the Taylor frozen-flow method, in which a Kolmogorov phase screen is repeatedly shifted according to some velocity to simulate the effect of wind in the atmosphere \cite{Schmidt, Srinath, Vogel, PoyneerExperimental}. This generalization, which we call ``boiling flow," accounts for random perturbations in the atmosphere by adding random Kolmogorov phase screens at each time-step. Boiling flow is a computationally efficient algorithm which generates physically-relevant phase screens and improves on pure frozen-flow \cite{Srinath}; as a result, it has become a common and well-known method for generating atmospheric phase screens (i.e., phase aberration data induced by atmospheric turbulence) \cite{Gerard, Dayton, Lin, PoyneerLaboratory, Jia, Boddeda, Sheikh, Snyder, Sridhar, Lehtonen, eBraga}. However, its use depends on the parameters $(L_0, r_0, v_x, v_y)$ and a new parameter, $\alpha$, which we call the \textit{boiling coefficient}. Many uses of boiling flow select these parameters based on a desired physical environment \cite{Gerard, Dayton, Jia, Boddeda, Sheikh, Lehtonen, eBraga}. However, because these parameters may not be physically well-defined for aero-optic effects, it is difficult to choose these parameters based on an aero-optic environment. Thus, instead of choosing the parameters based on a physical environment, estimating the parameters from measured aero-optic phase aberration data could allow boiling flow to generate aero-optically relevant phase screens.

There exist methods to estimate the boiling flow parameters $(L_0, r_0)$ from physical measurements, but these methods do not directly fit the spatial statistics of measured phase aberrations. The outer scale $L_0$ is interpreted as the upper bound of the size of turbulent eddies in the atmosphere and the Fried coherence length $r_0$ is defined as the diameter over which the mean-squared phase error is at most one radian \cite{Schmidt, Fried}. Both of these values parameterize the spatial power spectral density (PSD) of random Kolmogorov phase screens and thus determine the spatial statistics of the phase screens. Using a variety of physical parameters, these values can be derived from the refractive-index structure parameter $C_n^2$ \cite{Schmidt, Coulman}. However, when these physical values are unknown, the parameters $(L_0, r_0)$ are often estimated from the spatial covariance of angle-of-arrival (AA) measurements \cite{ZiadFrom, Schöck, Avila, Ziad} or the Zernike modes of measured phase aberration data \cite{Schöck, Winker, Ziad}. These estimation methods allow researchers to understand and analyze the physical environment of these experiments, but do not directly fit the spatial statistics of measured data. Thus, random Kolmogorov phase screens that are generated using these estimates may not have the same spatial statistics as the measured phase aberrations.

An existing peak fitter and layer identification algorithm can estimate the wind velocity components $(v_x,v_y)$ and boiling coefficient $\alpha$ from atmospheric phase aberration measurements, but this method may not generalize to aero-optic data. This algorithm estimates the wind velocity components $(v_x,v_y)$ by locating the peaks of the temporal power spectral density (TPSD) of the (spatial) Fourier modes of the measured phase aberrations and applying the Taylor frozen-flow hypothesis \cite{PoyneerExperimental}. The value of $\alpha$ is then fit using the width of the TPSD peak(s) \cite{Srinath, PoyneerLaboratory}. This estimation method can be used to analyze the physical structure of atmospheric turbulence, identify multiple layers of frozen-flow, and verify the frozen-flow hypothesis \cite{PoyneerExperimental}. However, it is difficult to apply this method to aero-optic phase aberration data. For example, the method depends on several pre-set values (e.g., frequency ranges to search through, tolerance values, likelihood ratio) \cite{PoyneerExperimental} which may need to be adjusted for different measured data sets; as a result, the method is not fully automated \cite{PoyneerLaboratory}. Furthermore, the algorithm presented by Poyneer \textit{et al.} \cite{PoyneerExperimental} only searches for TPSD peaks below 100 Hz, while the bandwidth of aero-optic effects may exceed this range \cite{WangPhysicsComputation, NightingalePhaseLocked, BurnsEstimation}. This restriction can be relaxed to find the TPSD peaks of aero-optic phase aberration data, but this would significantly increase the computational expense of the brute-force algorithm for identifying wind velocities from the TPSD peaks \cite{PoyneerExperimental}. Therefore, existing peak fitter and layer identification algorithms may not give accurate estimates of the parameters $(v_x,v_y,\alpha)$ from measured aero-optic phase aberrations.

In this paper, we introduce a method to estimate the parameters of boiling flow from turbulent boundary layer (TBL) data containing measured aero-optic phase aberrations. Because we simulate a boundary layer, we focus on estimating parameters for only a single wind layer; the measured (and synthetic) phase aberration data is then a time-series of single-channel images. Instead of interpreting the boiling flow parameters as physical quantities, we interpret them as parameters of the spatial and temporal statistics of the phase aberrations. We thus introduce an algorithm that estimates these parameters to fit the statistics of the measured data; we then generate phase screens using these parameters. This method is computationally efficient and the temporal power spectral density (TPSD) of the slopes of the synthetic phase screens (i.e., the deflection angle $\theta_x$ in the stream-wise direction of the flow) reasonably matches that of the measured phase aberrations from two TBL data sets \cite{Kemnetz}. However, the TPSD of the phase screens (themselves) only matches that of the measured phase aberrations at relatively large frequencies and the Kolmogorov spatial structure function of the phase screens does not match that of the measured phase aberrations from either TBL data set.

\section{BOILING FLOW}
\label{s: Boiling Flow}
In this section, we describe the method of boiling flow for generating phase screens for a single wind layer. Srinath \textit{et al.} \cite{Srinath} describe this method in detail, but we introduce notation here which emphasizes the five parameters of boiling flow: $\theta = (L_0, r_0, v_x, v_y, \alpha)$. These parameters include the outer scale $L_0$ (m), Fried coherence length $r_0$ (m), flow velocity components $(v_x, v_y)$ (pixels per time-step), and boiling coefficient $\alpha$.

Boiling flow generates a time-series of random Kolmogorov phase screens. A random Kolmogorov phase screen is an $N$-by-$N$ two-dimensional array $\phi \in \R^{N\times N}$ sampled from a probability distribution that is specified by an analytical (spatial) power spectral density (PSD); this analytical PSD is derived from the Kolmogorov theory of turbulence \cite{Schmidt}. Kolmogorov phase screens are generated by sampling from this probability distribution in the frequency domain; that is, the algorithm first samples the (spatial) Fourier modes and then applies an inverse Fourier transform. To write an analytical formula for this algorithm, we define the following notation:
\begin{itemize}
    \item $(f_x,f_y)$ are the frequency components in units if $m^{-1}$.    
    \item $\delta >0$ is the pixel spacing in units of meters.
    \item $(f_{x_k}, f_{y_\ell})$ are the frequency bins, in units of cycles per sample, for $0 \leq k,\ell< N$. That is, $\Bigl(\frac{f_{x_k}}{\delta},\frac{f_{y_\ell}}{\delta}\Bigr)$ has units of $m^{-1}$. As a result, $-\frac{1}{2}\leq f_{x_k}, f_{y_\ell} \leq \frac{1}{2}$ for all $k, \ell$.
    \item $\tilde{\phi}_n(f_{x_k}, f_{y_\ell})$ is the two-dimensional discrete-space Fourier transform of the phase screen at time $n$, $\phi_n\in\R^{N\times N}$, at frequency bins $(f_{x_k}, f_{y_\ell})$. That is, $\tilde{\phi}_n$ contains the Fourier modes.
    \item $w_n(f_{x_k}, f_{y_\ell})$ is complex-valued white noise. That is, the real and imaginary parts of $w_n(f_{x_k}, f_{y_\ell})$ are independent and both distributed as $\mathcal{N}(0,1)$.

    \item $S_\phi(f_x,f_y)$ is an assumed two-dimensional (spatial) PSD of each phase screen $\phi_n$ at frequencies $(f_x,f_y)$.
\end{itemize}
For the spatial PSD, boiling flow uses the the Von K\`arman PSD:
    \begin{align}\label{eq: Von Karman PSD}
        S_\phi(f_x,f_y) = 0.023 \times r_0^{-5/3} \times (f_x^2+f_y^2+L_0^{-2})^{-11/6}.
    \end{align}
Equation~(\ref{eq: Von Karman PSD}) depends on two parameters: the outer scale $L_0$ and the Fried coherence length $r_0$ \cite{Schmidt}. The Fourier modes of random Kolmogorov phase screens are then given by
\begin{align}\label{eq: Kolmogorov Phase Screen}
    \tilde{\phi}(f_{x_k}, f_{y_\ell}) = \frac{1}{N\delta} \:\sqrt{S_{\phi}\Biggl(\frac{f_{x_k}}{\delta},\frac{f_{y_\ell}}{\delta}\Biggr)}\hspace{0.2cm} w_n(f_{x_k}, f_{y_\ell}).
\end{align}
The phase screens $\phi\in\R^{N\times N}$ are obtained by taking the real part of the inverse fast Fourier transform (FFT) of $\tilde{\phi}$.

\subsection{Frozen-flow}
\label{s: Frozen-flow}
Boiling flow generates a time-series of phase screens using the Taylor frozen-flow method. This method translates a phase screen according to some velocity \cite{Schmidt, Srinath, Vogel, PoyneerExperimental}; applying this translation recursively (i.e., to the previous time-step) constructs a time-series of phase screens. Frozen-flow simulates a dynamically evolving turbulent medium, typically characterized by wind in the atmosphere \cite{Schmidt}. Given some velocity $(v_x, v_y)$ in units of pixels per time-step, frozen-flow translates the previous phase screen $\phi_{n-1}$ in the frequency domain:
\begin{align}\label{eq: Frozen-flow}
    \tilde{\phi}_n(f_{x_k}, f_{y_\ell}) = e^{-j2\pi (v_xf_{x_k}+v_yf_{y_\ell})}\tilde{\phi}_{n-1}(f_{x_k}, f_{y_\ell}).
\end{align}
Applying this translation to the Fourier modes allows the parameters of frozen-flow, $(v_x,v_y)$, to be non-integer values.

\subsection{Atmospheric Boiling}
\label{s: Atmospheric boiling}
Boiling flow then generalizes frozen-flow to account for atmospheric boiling. Atmospheric boiling refers to random phase perturbations in a turbulent medium that occur at each time-step \cite{Srinath}. To account for this phenomenon, boiling flow adds new (temporally independent) Kolmogorov phase screens at each time-step. For simplicity of notation, we define the scaling term for the variance of $w_n(f_x,f_y)$ in Kolmogorov phase screens:
\begin{align}
    P(f_{x_k}, f_{y_\ell}) = \frac{1}{N\delta}\sqrt{S_{\phi}\Biggl(\frac{f_{x_k}}{\delta},\frac{f_{y_\ell}}{\delta}\Biggr)}.
\end{align}
Boiling flow applies the following equation at each time-step $n$ to generate the Fourier modes of each phase screen:
\begin{align}\label{eq: Boiling Flow}
    \tilde{\phi}_n(f_{x_k}, f_{y_\ell}) = \alpha \:e^{-j2\pi (v_xf_{x_k}+v_yf_{y_\ell})}\:\tilde{\phi}_{n-1}(f_{x_k}, f_{y_\ell})+\sqrt{1-\alpha^2}\:P(f_{x_k}, f_{y_\ell})\:w_n(f_{x_k}, f_{y_\ell}).
\end{align}
The boiling coefficient $\alpha \in (0,1]$ controls the strength of the random phase fluctuations occurring at each time-step. The choice to weight the sum of the frozen-flow component and the random phase screen by $\alpha$ and $\sqrt{1-\alpha^2}$ (respectively) ensures that the spatial energy spectrum is preserved at each time-step. As a result, each phase screen $\phi_n$ is still a Kolmogorov phase screen (i.e., its spatial PSD is the Von K\`arman PSD).

To apply boiling flow, we use Eq.~(\ref{eq: Boiling Flow}) to generate a time-series of phase screens. We start with an initial Kolmogorov phase screen,
\begin{align}
    \tilde{\phi}_0 (f_{x_k}, f_{y_\ell})=P(f_{x_k}, f_{y_\ell})\:w_0(f_{x_k}, f_{y_\ell}),
\end{align}
and apply Eq.~(\ref{eq: Boiling Flow}) recursively for some $N_s$ time-steps. At each time-step, we take the real part of the inverse FFT of $\tilde{\phi}_n$ to recover a phase screen $\phi_n\in\R^{N\times N}$. We have added two additional steps to the algorithm to make the phase screens more physically relevant:
\begin{itemize}
    \item First, to prevent phase wrapping, we generate phase screens four times larger than the desired (square) aperture size. After taking inverse FFTs at each time-step, we restrict to the top left quadrant of the over-sized phase screens. This is necessary since our implementation uses FFTs, which cause the frequency shifts $v_xf_{x_k}+v_yf_{y_\ell}$ to ``wrap around" the domain of $-1/2\leq f_{x_k}, f_{y_\ell}\leq 1/2$. This phenomenon, called phase-wrapping, can cause periodicity in long time-series. However, our method of generating over-sized phase screens (and then restricting to a smaller region within each screen) avoids this periodicity.

    \item Second, after restricting to the appropriately-sized phase screens $\phi_n\in\R^{N\times N}$, we remove tilt, tip, and piston (TTP) from each phase screen. This is a common practice for post-processing the aero-optic measurements from an experiment \cite{Kemnetz}, so this step improves the physical relevance of the synthetic phase screens.
\end{itemize}
This method provides a way to generate time-series of phase screens with arbitrarily many time-steps.

\section{PARAMETER ESTIMATION FROM TRAINING DATA}
\label{s: Parameter Estimation from Training Data}
In this section, we outline our methodology for estimating the boiling flow parameters from a training data set of measured aero-optic phase aberrations. The training data is a time-series of $N_T$ single-channel images (with aperture size $M\times N$) whose pixel values are phase aberrations. Note that we allow the training data images to be rectangular instead of square. Thus, for the training calculations, we take the largest square we can inscribe inside of the aperture, with some length $K$: $\phi_n\in\R^{K\times K}$ for $n = 0, 1, \dots, N_T-1$. We now introduce a method to estimate $\theta = (L_0, r_0, v_x, v_y, \alpha)$ from the training data $\phi_0,\phi_1,\dots,\phi_{N_T-1}$.

\subsection{Boiling Parameters}
\label{s: Boiling parameters}
We first estimate the parameters $(L_0, r_0)$ based on the analytical form of $S_\phi(f_x,f_y)$. We call $(L_0, r_0)$ the \textit{boiling parameters} since they parameterize the spatial PSD of random Kolmogorov phase screens. Furthermore, because the form of Eq.~(\ref{eq: Boiling Flow}) preserves the spatial PSD at each time-step, $(L_0, r_0)$ parameterize the spatial statistics of each phase screen.

We set $\hat{L}_0$ to the length of the aperture of the training data. In practice, the parameter $L_0$ is often set to $\infty$, reducing Eq.~(\ref{eq: Von Karman PSD}) to the Kolmogorov PSD \cite{Schmidt}. However, this results in $S_\phi(0,0) = \infty$, which presents a numerical issue. Thus, we set $\hat{L}_0 = N\delta$ (meters), where $N$ is the length of the larger axis (in pixels) of the rectangular aperture. Although we estimate the remaining parameters of boiling flow to match the statistics of the training data, we use physical justification for this parameter: since the optical measurements are sensitive to only the aberrations which are at most the size of the aperture \cite{Roddier}, the measured data will only reflect the effects of turbulent eddies which are at most this size. Because $L_0$ is the upper bound of the size of turbulent eddies \cite{Schmidt}, $L_0$ values larger than the size of the aperture will not make a significant difference in the phase aberration measurements.

After computing $\hat{L}_0$, we estimate $r_0$ from the spatial PSD of the measured phase aberrations, $\hat{S}_\phi(f_x,f_y)$. Although the Fried coherence length $r_0$ has a physical definition for atmospheric turbulence, it is not well-defined for aero-optic effects. This follows from the fact that the definition of $r_0$ assumes that the spatial statistics of the phase aberrations are isotropic, which is not the case for aero-optic effects \cite{Vogel, Siegenthaler}. Thus, we consider $r_0$ only as a parameter that describes the spatial statistics of the training data. To estimate this value, we first estimate the spatial PSD of each training image (i.e., each time-step of the training data) by extending the well-known Welch's method \cite{Welch} to computation of a two-dimensional PSD. Because the measured data will often have low spatial resolution, we do not break the images into smaller blocks; instead, we take the full square images $\phi_n\in\R^{K\times K}$. Our methodology is then as follows:
\begin{enumerate}
    \item First, we remove the spatial mean from each image. This step is necessary to remove spectral leakage in low frequencies \cite{Oppenheim}, especially since Welch's method assumes that the signal is zero-mean \cite{Welch}.

    \item We then construct a two-dimensional Hamming window $H = h h^T \in \R^{K\times K}$, where $h\in\R^K$ is a (standard one-dimensional) Hamming window of length $K$. We multiply each image by $H$ (element-wise).

    \item Next, we compute the magnitude-squared (two-dimensional) FFT of each image and divide by the sum of the squared Hamming window coefficients (i.e., the Welch scaling factor) \cite{Welch, Oppenheim}.

    \item We then divide the resulting spatial PSD values by the squared spatial sampling frequency $1/\delta^2$ (which has units of 1/$m^2$) to convert the PSD units from energy per sample to energy per $m^2$ \cite{Oppenheim}.

    \item Finally, we average the spatial PSD estimates of each image.
\end{enumerate}
After estimating $\hat{S}_\phi(f_x,f_y)$ from the training data using the above methodology, we plug $\hat{L}_0$ into Eq.~(\ref{eq: Von Karman PSD}) and solve for $r_0$. This results in a function
\begin{align}\label{eq: r_0 Estimation}
    \hat{r}(f_x, f_y) = \Biggl(\frac{0.023\times(f_x^2+f_y^2+\hat{L}_0^{-2})^{-11/6}}{\hat{S}_\phi(f_x,f_y)}\Biggr)^{3/5}.
\end{align}
In theory, $\hat{r}(f_x, f_y)$ is constant across various $(f_x,f_y)$ values. However, this is likely not the case for aero-optic phase aberrations; there will be fluctuations, especially at the high frequencies. Thus, we average the values $\hat{r}(f_x, f_y)$ over a set of frequency bins $(f_x,f_y)$ to compute $\hat{r}_0$. We remove the $(0,0)$ frequency and all pairs $(f_x,f_y)$ containing the largest three $f_x$ or $f_y$ bins from this average to reduce noise.

\subsection{Flow Parameters}
\label{s: Flow parameters}
Once the values $(\hat{L}_0, \hat{r}_0)$ are computed, we estimate the velocity components $(v_x,v_y)$ and boiling coefficient $\alpha$. We call $(v_x, v_y, \alpha)$ the \textit{flow parameters} since they parameterize both the flow velocity and the fraction of each phase screen that is computed using pure frozen-flow. Furthermore, $(v_x,v_y,\alpha)$ parameterize the temporal statistics of the training data.

We determine $(\hat{v}_x, \hat{v}_y)$ by locating the pixel shift which maximizes the spatial cross-correlation of the training data. The cross-correlation of a time-series of discrete-space 2-D arrays $\phi_n(x_i,y_j)$ with lag $L$ is given by
\begin{align}\label{eq: Cross-Correlation}
    R(i,j; L) = \E\bigl[\phi_n(x_r,y_s)\;\phi_{n-L}(x_{r-i},y_{s-j})\bigr],
\end{align}
where $(r,s)$ varies across the (finite) 2-D spatial grid. This value $R(i,j;L)$ is the correlation between pixel pairs at a distance $(i,j)$, which we call the \textit{pixel shift}, after $L$ time-steps. We estimate Eq.~(\ref{eq: Cross-Correlation}) from the experimental data via
\begin{align}\label{eq: Cross-Correlation Estimate}
    \hat{R}(i,j; L) = \frac{1}{N_T-L}\sum_{n=L}^{N_T-1}\sum_{(r,s)}\phi_n(x_r,y_s)\phi_{n-L}(x_{r-i},y_{s-j}).
\end{align}
We then identify the shifts $(\hat{i}, \hat{j})$ which result in the highest correlation:
\begin{align}\label{eq: Max Cross-Correlation}
    (\hat{i}, \hat{j}) = \underset{(i,j)}{\text{argmax}}\;\hat{R}(i,j;L).
\end{align}
To refine this estimate, we use parabolic interpolation. That is, we fit two parabolas to values of $\hat{R}(k,\ell; L)$: one for $k =\hat{i}-1,\hat{i},\hat{i}+1$ and $\ell = \hat{j}$, the other for $\ell=\hat{j}-1,\hat{j},\hat{j}+1$ and $k = \hat{i}$. We then find the maximal indices of each parabola, $k=i^*$ and $\ell=j^*$. The velocity estimates, in units of pixels per time-step, are then given by
\begin{align}\label{eq: Velocity Esimates}
    (\hat{v}_x, \hat{v}_y) = \frac{1}{L}(i^*,j^*).
\end{align}
Here, we choose $L=10$ since this value is large enough to give a precise estimate of $(\hat{v}_x, \hat{v}_y)$ and small enough that Eq.~(\ref{eq: Cross-Correlation Estimate}) does not miss the correlations imposed by the flow.

After estimating $(\hat{v}_x, \hat{v}_y)$ from the training data, we compute $\hat{\alpha}$ using least-squares regression. That is, we find
\begin{align}\label{eq: Boiling Flow alpha Problem}
    \hat{\alpha} = \underset{\alpha\in\R}{\text{argmin}}\Biggl\{\sum_{n=1}^{N_T-1}\sum_{k,\ell=0}^{K-1} \Big\vert\tilde{\phi}_n(f_{x_k}, f_{y_\ell}) - \alpha \;e^{-j2\pi(\hat{v}_xf_{x_k}+\hat{v}_yf_{y_\ell})}\tilde{\phi}_{n-1}(f_{x_k}, f_{y_\ell}) \Big\vert^2\Biggr\}.
\end{align}
This value is the correlation coefficient between the shifted phase screen from the previous time-step and the phase screen at the current time step. We arrive at a closed-form solution to Eq.~(\ref{eq: Boiling Flow alpha Problem}) using matrix calculus and compute $\hat{\alpha}$ with vector inner-products. We compute $\hat{\alpha}$ using the Fourier modes (instead of the phase values) to accommodate for non-integer velocity components $(\hat{v}_x, \hat{v}_y)$.

\section{RESULTS}
\label{s: Results}
In this section, we show results from phase screens generated by boiling flow with the parameter estimation algorithm described in Sec.~\ref{s: Parameter Estimation from Training Data}. We evaluate this method on the basis of how closely the synthetic phase screens match relevant spatial and temporal statistics of the measured data. Our results indicate that the synthetic phase screens reasonably match the temporal statistics of the slopes of the measured data. However, the phase screens only match the relatively short-range temporal statistics of the measured data itself and do not match the spatial statistics of the measured data.

\subsection{Methodology}
\label{s: Methodology}
We trained and tested boiling flow on two turbulent boundary layer (TBL) data sets containing measured aero-optic phase aberrations \cite{Kemnetz}. Section~\ref{s: Experimental data} describes the experimental setup used to measure the TBL data; the resulting data sets contain time-series of single-channel images whose pixel values are phase aberrations. We trained boiling flow (for a single wind layer) on the first 80\% of each time-series and compared the synthetic phase screens to the remaining 20\% of the measured time-series (i.e., validation data). We used the method in Sec.~\ref{s: Parameter Estimation from Training Data} to train boiling flow and then applied the method of Sec.~\ref{s: Boiling Flow} to generate phase screens. Section~\ref{s: Quality metrics} describes the statistics used to evaluate this method. We averaged the statistics over twenty synthetic time-series, each with twenty times the number of time-steps of the validation data, to reduce noise.

\subsubsection{Experimental data}
\label{s: Experimental data}
Kemnetz and Gordeyev \cite{Kemnetz} measured the TBL data through a high-speed wind tunnel experiment. The authors held free stream velocity in the wind tunnel constant at $M =0.4$, which induced a TBL with estimated thickness $\delta=0.0156$ m. They then propagated a $\lambda = 532$ nm wavelength laser through the TBL and took measurements using a Shack Hartmann Wavefront Sensor (SHWFS). We note that the data from this experiment is highly convective and fairly homogeneous. Thus this data provides a reasonable physical structure for boiling flow to simulate.

The authors computed phase aberration data from the SHWFS measurements. A SHWFS measures the deflection angles
\begin{align}\label{eq: Deflection Angles}
    \Theta(x,y,t) = \bigl(\theta_x(x,y,t), \theta_y(x,y,t)\bigr)
\end{align}
of detected light waves at times $t$ and grid locations $(x,y)$ on the sensor's aperture plane \cite{Kemnetz, Holmes, Freeman}. Deflection angle is related to phase aberrations $\phi(x,y,t)$ via the following relation \cite{Kemnetz, SiegenthalerShear, Cress}:
\begin{align}\label{eq: Phase theta Relationship}
    \nabla_{(x,y)}\bigl\{\phi(x,y,t)\bigr\} = \frac{2\pi}{\lambda}\Theta(x,y,t).
\end{align}
Thus, Kemnetz and Gordeyev solved Eq.~(\ref{eq: Phase theta Relationship}) for $\phi(x,y,t)$ at each time $t$ using the deflection angle data. The resulting phase aberration data was then post-processed to remove TTP \cite{Kemnetz}.

Table~\ref{tab: Experimental Data Sets} lists details of the resulting aero-optic time-series data. Pixels on the edge of the images were removed to reduce noise, so the total number of pixels is less than the area of the aperture. To apply boiling flow, we set $N$ (the length of the square phase screens) as the maximum of the two axis lengths of the rectangular aperture. After generating phase screens, we reduced to the aperture size of the measured data and removed the edge points.

\begin{table}[htbp]
    \centering
    \caption{Experimental Data Sets}
    \begin{tabular}{|c||c|c|}
        \hline 
        \large \textbf{Data Set} & \large F06 & \large F12 \\
        \hline
        \textbf{Aperture Size (pixels)} & $25\times 34$ & $21\times 22$ \\
        \textbf{Number of Pixels} & 735 & 380 \\
        \textbf{Pixel Spacing $\delta$ (mm)} & 2.24 & 2.24 \\
        \textbf{Number of Time-Steps} & 150,600 & 251,100 \\
        \textbf{Sampling Frequency (Hz)} & 100,000 & 130,000 \\
        \hline 
    \end{tabular}
    \label{tab: Experimental Data Sets}
\end{table}

\subsubsection{Quality metrics}
\label{s: Quality metrics}
Our quality metrics evaluate this method's accuracy by comparing the TPSD and Kolmogorov spatial structure function of the synthetic phase screens with those of the measured data. The most important quality metric is the error between the TPSD of the slopes of the synthetic phase screens and the TPSD of the slopes of the measured data, followed by the error between the TPSD of the phase screens and the measured data (themselves). Lastly, we evaluate the error between the Kolmogorov spatial structure functions of the measured and synthetic data.

The TPSD of a time-series of data computes a quantity proportional to power (energy/sec) as a function of temporal frequency (Hz). Comparing the TPSD of measured data with the TPSD of synthetic data allows us to assess how well our method matches the temporal statistics of the measured data. Further, because a SHWFS directly measures deflection angles and since the $x$-axis is the stream-wise direction of the flow field \cite{Kemnetz}, we are most interested in matching the TPSD of $\theta_x$. We compute the time-series of $\theta_x$ images from the measured phase aberration data and synthetic phase screens by approximating the derivative with respect to the $x$-axis using two-sided finite differences.

We estimated the TPSD of both $\phi$ and $\theta_x$ by averaging the TPSD of each pixel's time-series. We used the well-known Welch's method \cite{Welch} to estimate the TPSD of each pixel. That is, we performed the following steps:
\begin{enumerate}
    \item First, we removed the temporal mean from each image pixel. This step is necessary to remove spectral leakage in low frequencies \cite{Oppenheim}, especially since Welch's method assumes that the signal is zero-mean \cite{Welch}. Importantly, this step is performed on the entire signal and not on each time-block to avoid underestimating the power at low frequencies \cite{PoyneerExperimental}.
    
    \item Next, we broke up the time-series into smaller time-blocks of some length $N_b$ with 50\% overlap (i.e., consecutive blocks share $N_b/2$ time-steps). For data set F06, we used $N_b=596$ and $N_b=298$ time-steps for the TPSD of $\phi$ and $\theta_x$, respectively. For data set F12, we used $N_b=994$ for $\phi$ and $N_b=496$ for $\theta_x$.

    \item We then applied a Hamming window to each time-block.

    \item Next, we computed the magnitude-squared FFT of each mean-removed block and divided by the sum of the squares of each Hamming window coefficient (i.e., the Welch scaling factor) \cite{Welch, Oppenheim}.

    \item We then divided the resulting TPSD values by the temporal sampling frequency $f_s$ to convert the TPSD from units of energy per sample to energy per second \cite{Oppenheim}.

    \item Finally, we averaged the TPSD estimates over all time-blocks.
\end{enumerate}
After using this procedure to estimate the TPSD of each pixel, we averaged these estimates over the entire aperture.

Lastly, we use the Kolmogorov spatial structure function to evaluate the spatial statistics of the measured and synthetic data. The Kolmogorov spatial structure function is given by
\begin{align}\label{eq: Spatial Structure Function}
    D_\phi(\x_1, \x_2) = \E\Bigl[ \bigl(\phi(\x_1) - \phi(\x_2)\bigr)^2\Bigr],
\end{align}
where $\x = (x,y)$ is a two-dimensional grid location. For phase errors that are spatially homogeneous and isotropic (e.g., phase errors induced by atmospheric turbulence \cite{Vogel, Siegenthaler}), $D_\phi$ depends only on the \textit{separation distance} $r = \|\x_2 - \x_1\|$. However, aero-optic effects have been shown to be anisotropic \cite{Vogel, Siegenthaler}. Thus, we generalize the structure function for aero-optic effects.

To get useful information from the structure function of aero-optic phase aberrations, we scale the data by its temporal standard deviation. That is, we first estimate
\begin{align}\label{eq: Mean Estimate}
    \hat{\mu}(\x) &= \frac{1}{N_T}\sum_{n=0}^{N_T-1} \phi_n(\x), \\
    \label{eq: Standard Deviation Estimate}
    \hat{\sigma}(\x) &= \sqrt{\frac{1}{N_T}\sum_{n=0}^{N_T-1}\bigl(\phi_n(\x)- \hat{\mu}(\x)\bigr)^2}
\end{align}
from the phase aberrations and then compute the structure function of $\phi/\hat{\sigma}$. Vogel \textit{et al.} \cite{Vogel} used this method to compensate for the heterogeneous spatial statistics of aero-optic effects, calling the resulting quantity the \textit{quasi-homogeneous} structure function
\begin{align}\label{eq: Quasi-Homogeneous Structure Function}
    D_{\phi/\sigma}(\x_1,\x_2) &= \E\Biggl[\Bigl(\frac{\phi(\x_1)}{\hat{\sigma}(\x_1)} - \frac{\phi(\x_2)}{\hat{\sigma}(\x_2)}\Bigr)^2\Biggr] \\ \label{eq: Equivalent Form of Structure Function}
    &= 2\Biggl(1 - \frac{\E\bigl[\phi(\x_1)\:\phi(\x_2)\bigr]}{\hat{\sigma}(\x_1)\hat{\sigma}(\x_2)}\Biggr).
\end{align}
This scaling ensures that the variance of the phase error at each grid location $\x$ is exactly one. Further, we see from Eq.~(\ref{eq: Equivalent Form of Structure Function}) that $D_{\phi/\sigma}$ evaluates the correlation coefficient between $\phi(\x_1)$ and $\phi(\x_2)$.

Because the scaled data is still spatially anisotropic, we compute $D_{\phi/\sigma}$ as a function of the separation distance $r$ and the angle $\theta$ between two points. That is, we introduce the \textit{anisotropic structure function}
\begin{align}\label{eq: Anisotropic Structure Function}
    \bm{D}_{\phi/\sigma}(r,\theta) = \text{mean}\Bigl\{D_{\phi/\sigma}(\x_1,\x_2):\;\|\x_2-\x_1\|=r, \text{ arg}(\x_2-\x_1)=\theta \text{ (mod }\pi\text{)}\Bigr\}.
\end{align}
To compute Eq.~(\ref{eq: Anisotropic Structure Function}), we first estimated Eq.~(\ref{eq: Quasi-Homogeneous Structure Function}) using a time-average and then computed Eq.~(\ref{eq: Anisotropic Structure Function}) directly from the $D_{\phi/\sigma}(\x_1,\x_2)$ data. Note that we explicitly computed the structure function in rectangular coordinates, $\bm{D}_{\phi/\sigma}(x,y)$, where $(x,y)$ is the two-dimensional difference (or separation) between two grid locations. However, because both $r$ and $\theta$ are inputs to the anisotropic structure function, this is equivalent to computing Eq.~(\ref{eq: Anisotropic Structure Function}).

Finally, we take the normalized root-mean squared error (NRMSE) between both the TPSD and structure functions of the synthetic phase screens and those of the measured data. To normalize the root-mean square error (RMSE), we divide the RMSE by the range of values between the 5th and 95th percentiles. Often, the normalized RMSE divides by the difference between the minimum and maximum values (i.e., the range of values), but this is sensitive to outliers in the data. We call our method \textit{stable range normalization} since taking the percentiles is less sensitive to outliers. First, let $\hat{S}_{\theta_x}$ and $\hat{S}_{\phi}$ denote the TPSD of $\theta_x$ and $\phi$ from the synthetic phase screens (while $S_{\theta_x}$ and $S_\phi$ denote those of the measured data). The TPSD scalar metrics are then
\begin{align}\label{eq: NRMSE of TPSD of Deflection Angle}
    NRMSE(S_{\theta_x},\hat{S}_{\theta_x}) &= \frac{\sqrt{\frac{1}{N_b}\sum_{i=0}^{N_b-1}\bigl(S_{\theta_x}(f_i) - \hat{S}_{\theta_x}(f_i)\bigr)^2}}{P_{95}(S_{\theta_x})-P_{5}(S_{\theta_x})} ,\\ \label{eq: NRMSE of the TPSD of Phase}
    NRMSE(S_{\phi}, \hat{S}_{\phi}) &= \frac{\sqrt{\frac{1}{N_b}\sum_{i=0}^{N_b-1}\bigl(S_{\phi}(f_i) - \hat{S}_{\phi}(f_i)\bigr)^2}}{P_{95}(S_{\phi})-P_{5}(S_{\phi})},
\end{align}
where $f_i$ are the frequency bins and $P_{95}(\cdot)$ and $P_5(\cdot)$ denote the 95th and 5th percentiles of an array of values (respectively). To evaluate the error between the structure functions of the measured and synthetic data, we take the square root of the values, $\bm{D}^{1/2}_{\phi/\sigma}(x,y) = \sqrt{\bm{D}_{\phi/\sigma}(x,y)}$; this places a larger weight on the smaller values of the structure function. As with the TPSD, we let $\bm{D}^{1/2}_{\phi/\sigma}$ denote the structure function of the measured data and $\bm{\hat{D}}^{1/2}_{\phi/\sigma}$ denote the structure function of the synthetic data. We then take the RMS over both axes,
\begin{align}\label{eq: NRMSE of Structure Function}
    NRMSE(\bm{D}^{1/2}_{\phi/\sigma}, \bm{\hat{D}}^{1/2}_{\phi/\sigma}) &= \frac{\sqrt{\frac{1}{N_x N_y}\sum_{k=0}^{N_y-1}\sum_{j=0}^{N_x-1}\bigl(\bm{D}^{1/2}_{\phi/\sigma}(x_j, y_k) - \bm{\hat{D}}^{1/2}_{\phi/\sigma}(x_j, y_k)\bigr)^2}}{P_{95}(\bm{D}^{1/2}_{\phi/\sigma})-P_{5}(\bm{D}^{1/2}_{\phi/\sigma})},
\end{align}
where $(x_k, y_j)$ denote the two-dimensional separations.

\subsection{Data simulation results}
\label{s: Data simulation results}
Table~\ref{tab: Parameter Estimates} lists the parameter estimates for each TBL data set computed using the method in Sec.~\ref{s: Parameter Estimation from Training Data}. Although the value of $\hat{L}_0$ is set based on physical characteristics of the experiment (i.e., the length of the aperture), the value of $\hat{r}_0$ likely has little physical relevance since it is not physically well-defined for aero-optic effects. However, the velocity components $(\hat{v}_x, \hat{v}_y)$ describe the convective nature of the data. Specifically, since the flow is streamwise along the $x$-axis, $\hat{v}_y$ is close to zero for both data sets. Lastly, although $\hat{\alpha}$ is not based on physical phenomena, the values we computed (above 0.91 for both data sets) validate the convective nature of the data: the phase screen at each time-step is highly correlated to the shifted phase screen from the previous time-step.

\begin{table}[htbp]
    \centering
    \caption{Parameter Estimates from Measured Data}
    \label{tab: Parameter Estimates}
    \begin{tabular}{|c||c|c|c|c|}
        \hline
        Data Set & $\hat{L}_0$ (m) & $\hat{r}_0$ (m) & $(\hat{v}_x, \hat{v}_y)$ (pixels per time-step) & $\hat{\alpha}$ \\
        \hline 
        F06 & 0.07616 & 0.20049 & (1.11919, -0.01313) & 0.91126 \\
        F12 & 0.04928 & 0.13077 & (0.519192, -0.00707) & 0.93024 \\
        \hline
    \end{tabular}
\end{table}

\begin{figure}[htbp]
    \centering
    \begin{subfigure}{\textwidth}
        \centering
        \includegraphics[width=\linewidth]{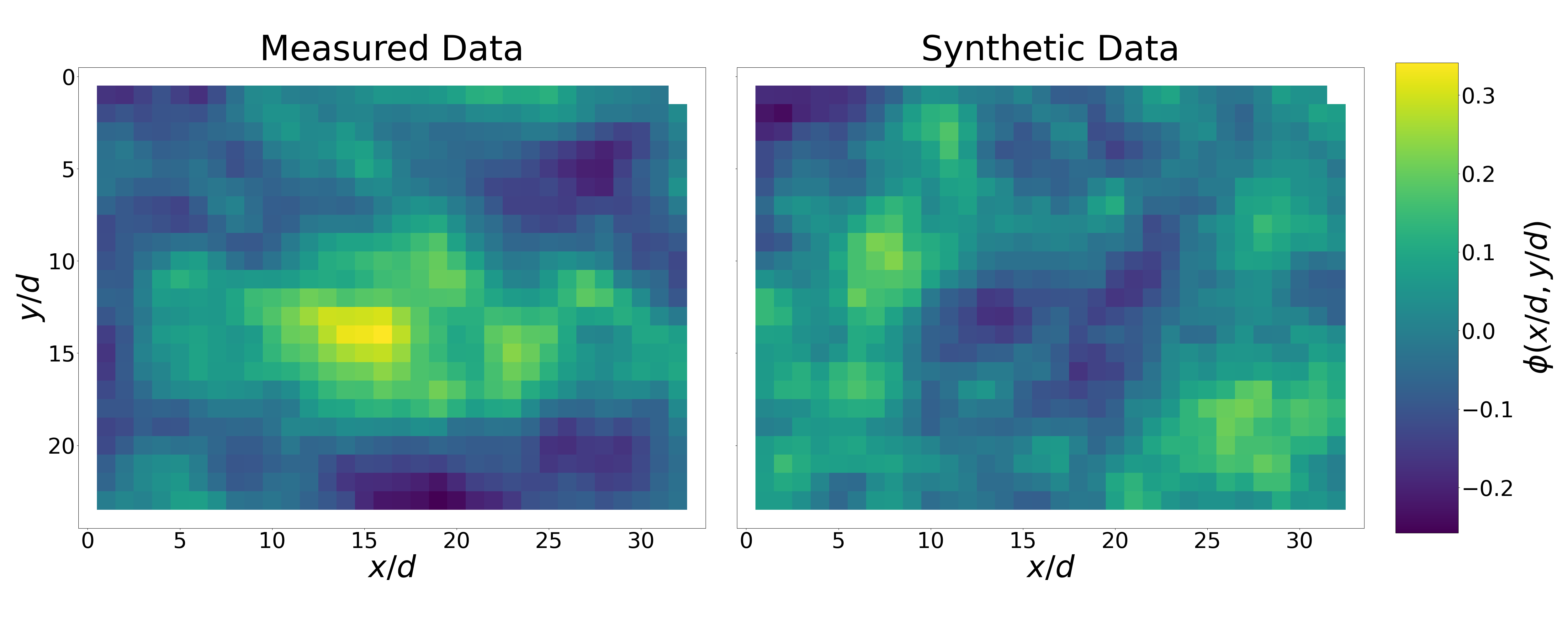}
        \caption{Data Set F06}
        \label{fig: F06 Image}
    \end{subfigure}
        \begin{subfigure}{\textwidth}
        \centering
        \includegraphics[width=\linewidth]{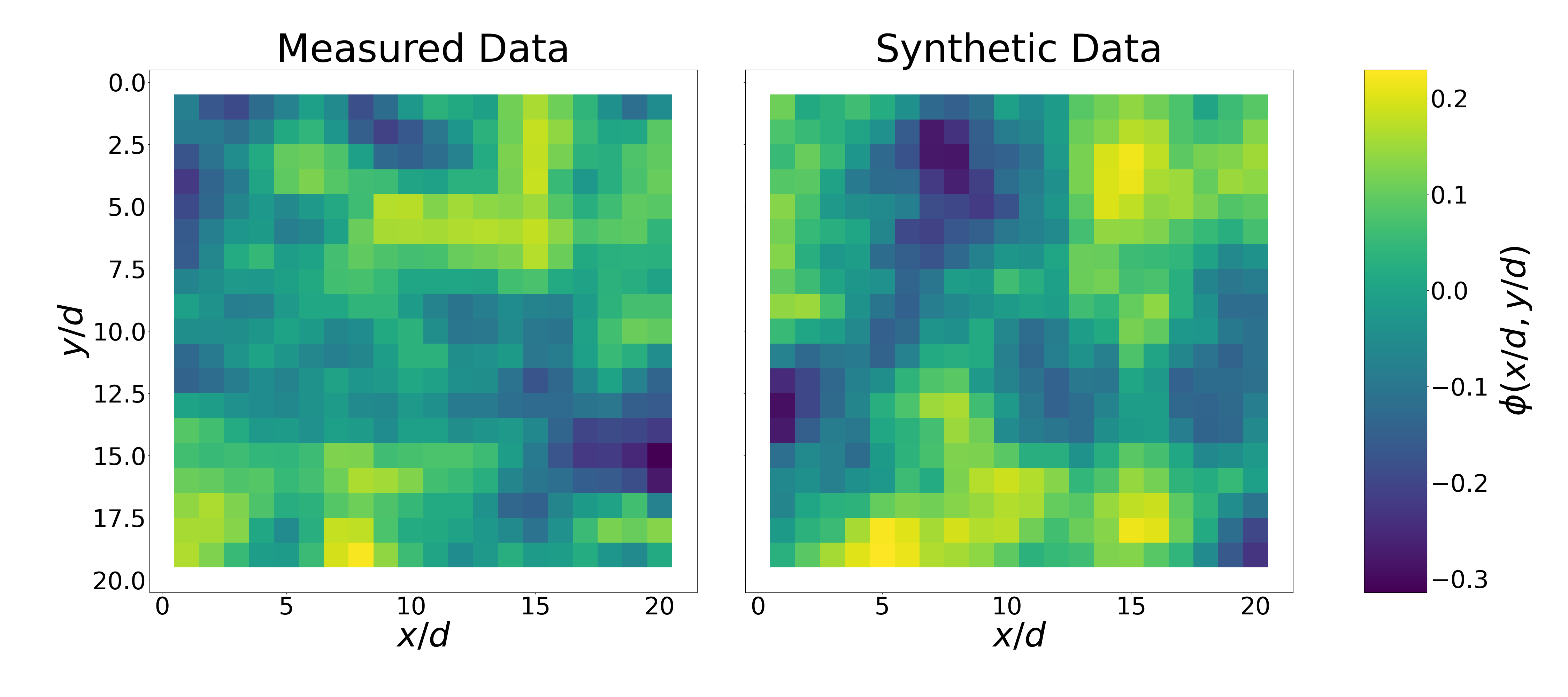}
        \caption{Data Set F12}
        \label{fig: F12 Image}
    \end{subfigure}
    \caption{This figure shows side-by-side images from the measured phase aberration data (left) and synthetic phase screens (right). Figure~\ref{fig: F06 Image} shows images for data set F06 and Fig.~\ref{fig: F12 Image} shows images for F12. The x- and y-axes show $x/d$ and $y/d$ (respectively), where $d$ is a sub-aperture spacing; $x/d$ and $y/d$ indicate the pixel index along each axis. The pixel values in each image are the phase values $\phi(x/d,y/d)$, quantized according to the color-bar on the right of each image pair. These images illustrate that boiling flow can generate phase screens on the same scale as the measured aero-optic phase aberrations and the images show close resemblance.}
    \label{fig: Images}
\end{figure}

\begin{figure}[htbp]
    \centering
    \begin{subfigure}{\textwidth}
        \centering
        \includegraphics[width=\linewidth]{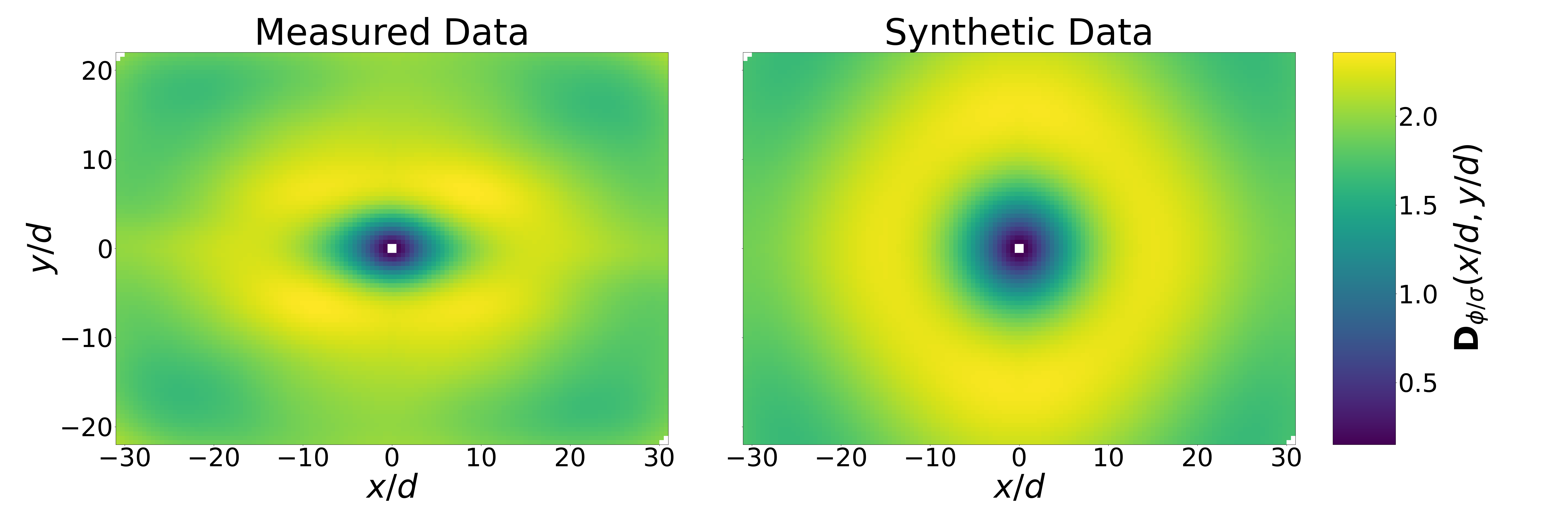}
        \caption{Data Set F06}
        \label{fig: F06 Structure Function}
    \end{subfigure}
        \begin{subfigure}{\textwidth}
        \centering
        \includegraphics[width=\linewidth]{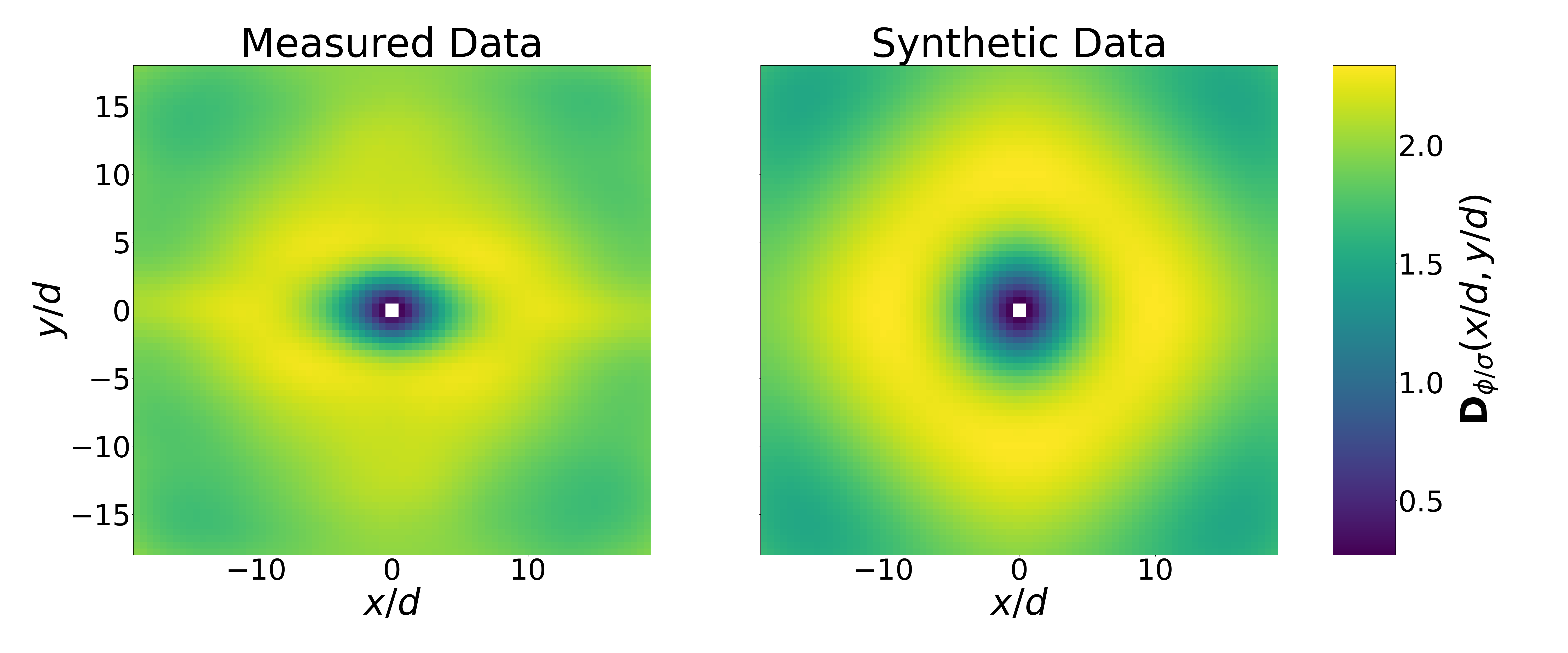}
        \caption{Data Set F12}
        \label{fig: F12 Structure Function}
    \end{subfigure}
    \caption{This figure shows images of the anisotropic structure function of the measured phase aberration data (left) and synthetic phase screens (right). Figure~\ref{fig: F06 Structure Function} shows the results for data set F06 and Fig.~\ref{fig: F12 Structure Function} shows the results for F12. The inputs to the structure function are $(x/d, y/d)$, where $d$ is the sub-aperture spacing; these input values are then the pixel distances between two grid locations. The structure function values are interpolated (using bi-linear interpolation) at the center of each pixel; the images are quantized according the color-bars on the right of each image. These results show that the contours of the structure function of the phase screens are circles, while the contours of the measured data's structure function are ellipses. Therefore, boiling flow does not accurately match the spatial statistics of measured aero-optic phase aberrations.}
    \label{fig: Structure Function}
\end{figure}

Figures~\ref{fig: Images} and~\ref{fig: Structure Function} show images of the phase screens and anisotropic structure function, respectively. The images in Fig.~\ref{fig: Images}, which show both the measured phase aberrations and (synthetic) phase screens, demonstrate that boiling flow can generate phase screens on the same scale as the measured phase aberrations. However, Fig.~\ref{fig: Structure Function} shows that the contours of the structure function of the synthetic phase screens are circles, while the contours of the measured data's structure function are ellipses. These two results follow from the Kolmogorov theory of turbulence's assumption of isotropic spatial statistics and the anisotropic spatial statistics of aero-optic phase aberrations \cite{Vogel, Siegenthaler}, respectively. The major axis of the ellipses for both measured data sets are along the x-axis since this is the streamwise direction of the flow; specifically, correlation induced by the flow remains strong at larger separation distances along the streamwise direction. Since boiling flow does not accurately fit these anisotropic characteristics, its model of the spatial statistics is not suitable for generating aero-optic phase screens.

\begin{figure}[htbp]
    \centering
    \begin{subfigure}{0.45\textwidth}
        \centering
        \includegraphics[width=\linewidth]{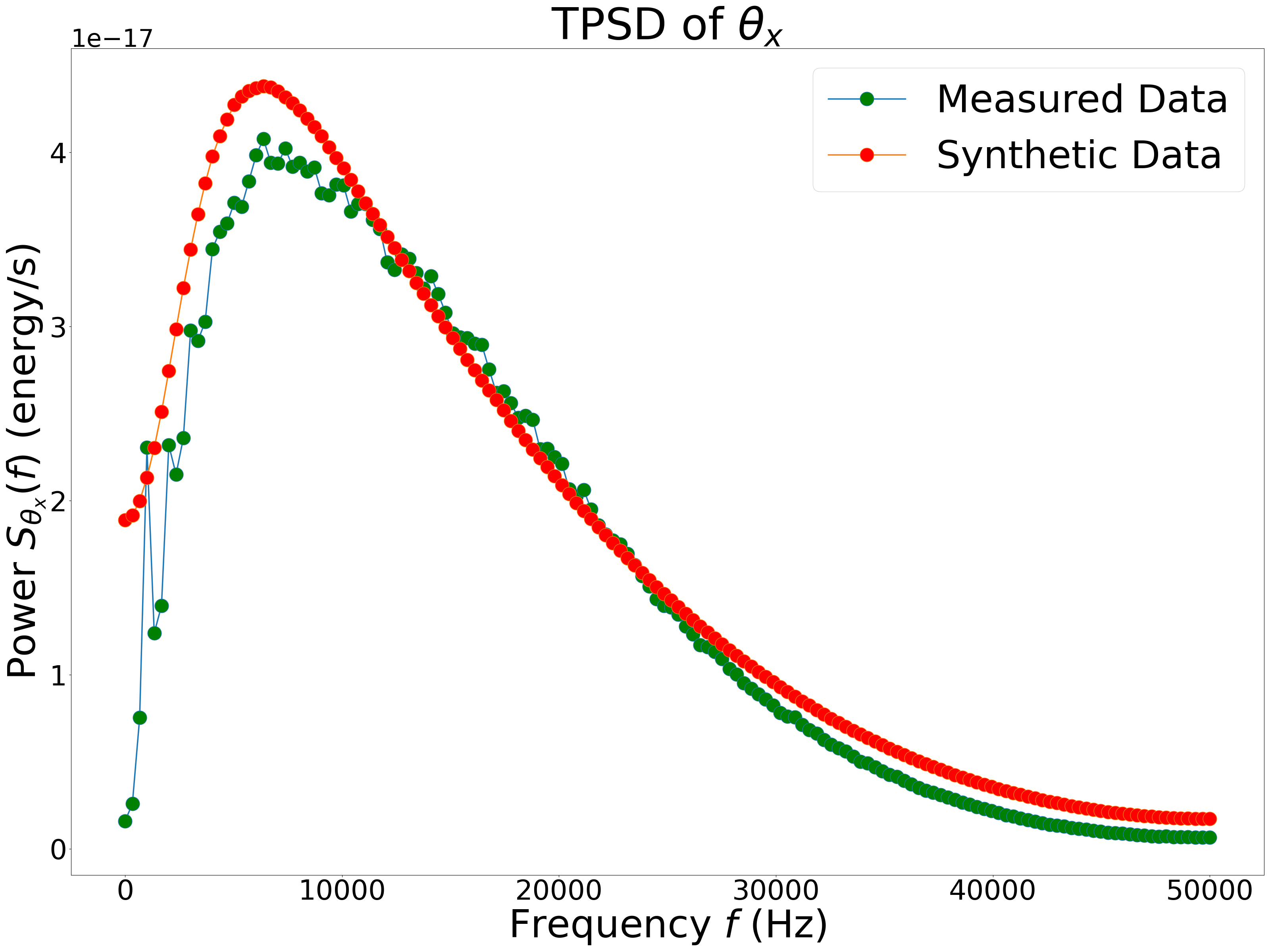}
        \caption{Data Set F06: TPSD of $\theta_x$}
        \label{fig: F06 TPSD of Deflection Angle}
    \end{subfigure}
    \begin{subfigure}{0.45\textwidth}
        \centering
        \includegraphics[width=\linewidth]{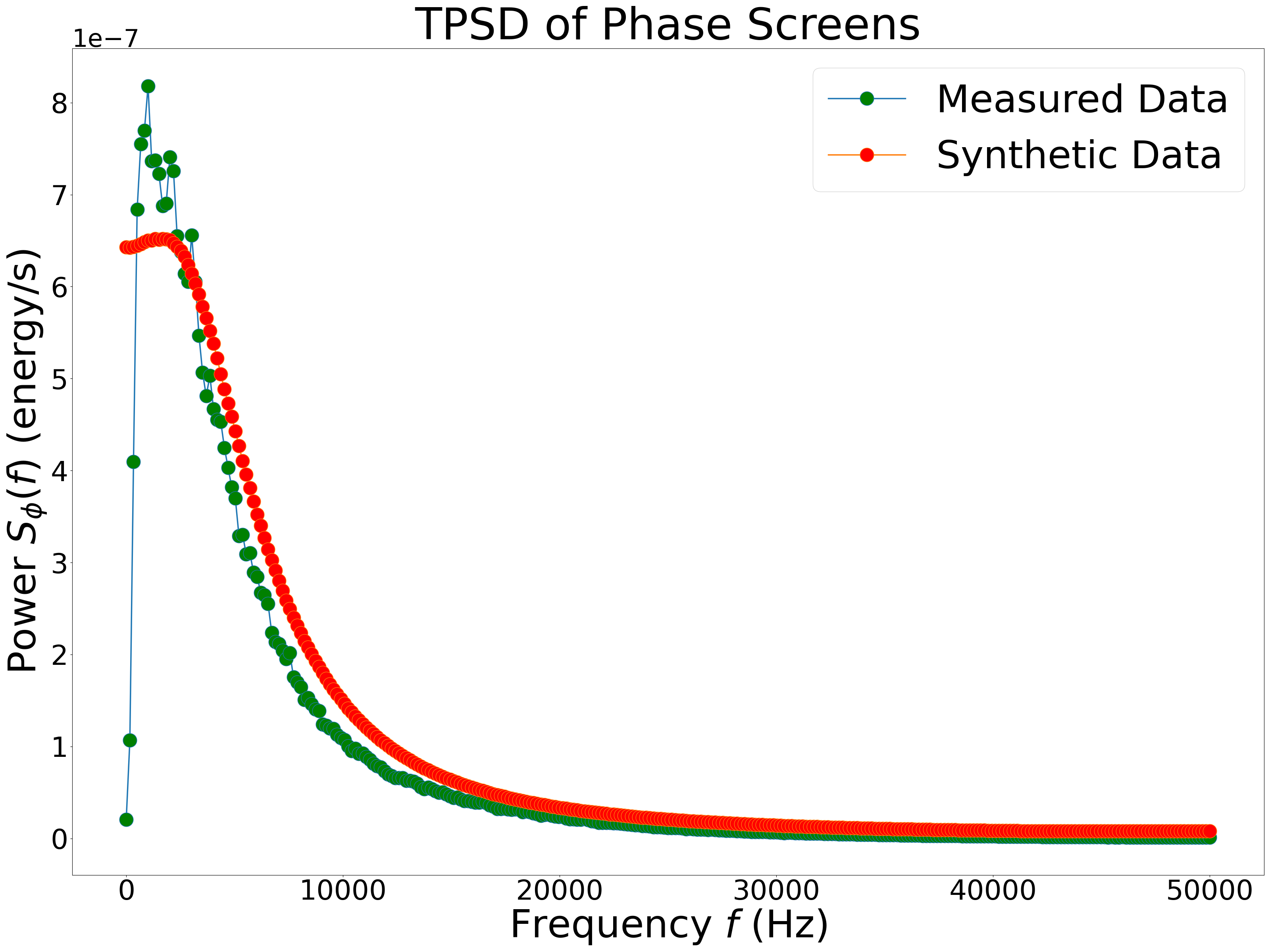}
        \caption{Data Set F06: TPSD of $\phi$}
        \label{fig: F06 TPSD of Phase}
    \end{subfigure}
    \begin{subfigure}{0.45\textwidth}
        \centering
        \includegraphics[width=\linewidth]{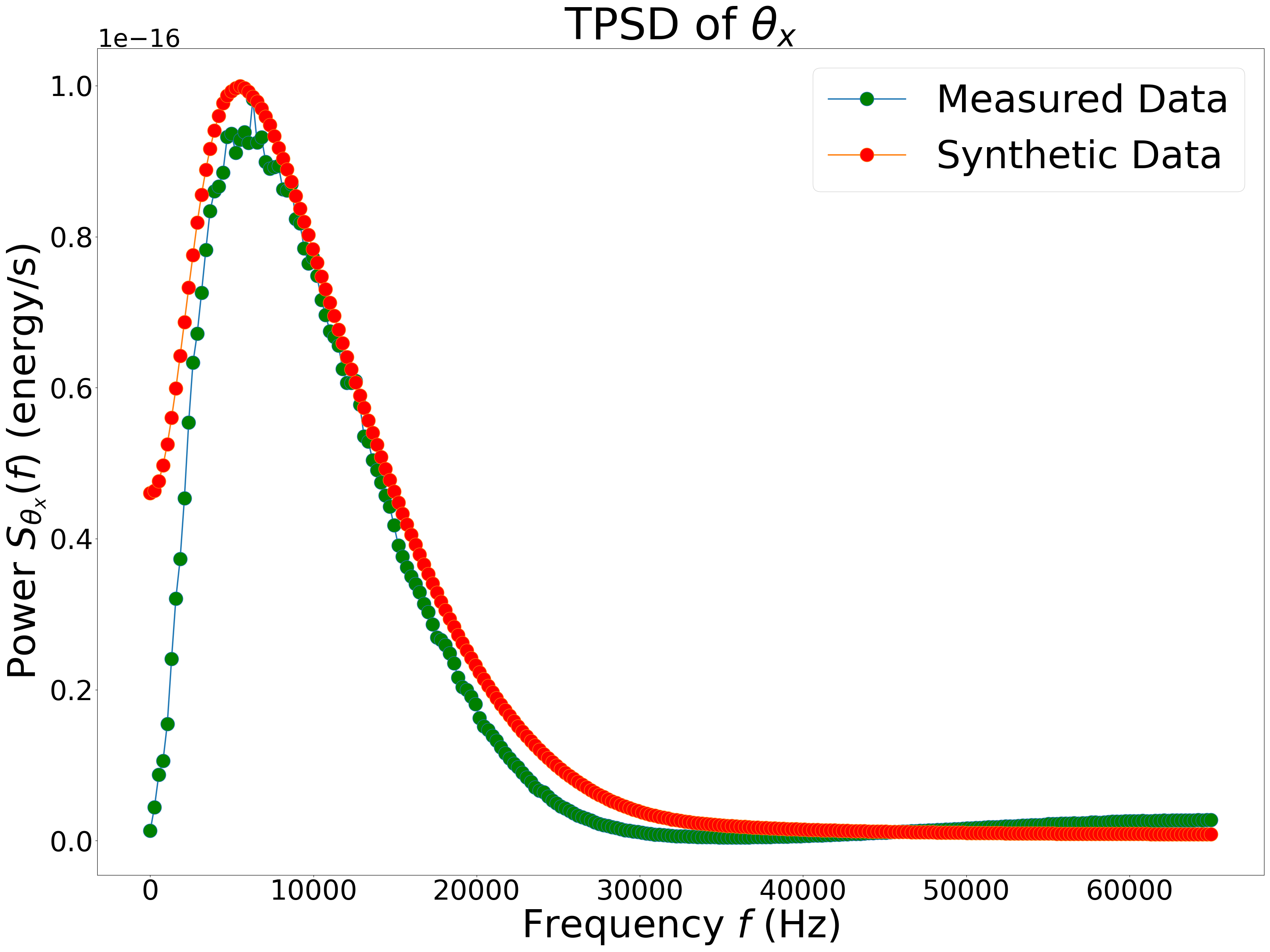}
        \caption{Data Set F12: TPSD of $\theta_x$}
        \label{fig: F12 TPSD of Deflection Angle}
    \end{subfigure}
    \begin{subfigure}{0.45\textwidth}
        \centering
        \includegraphics[width=\linewidth]{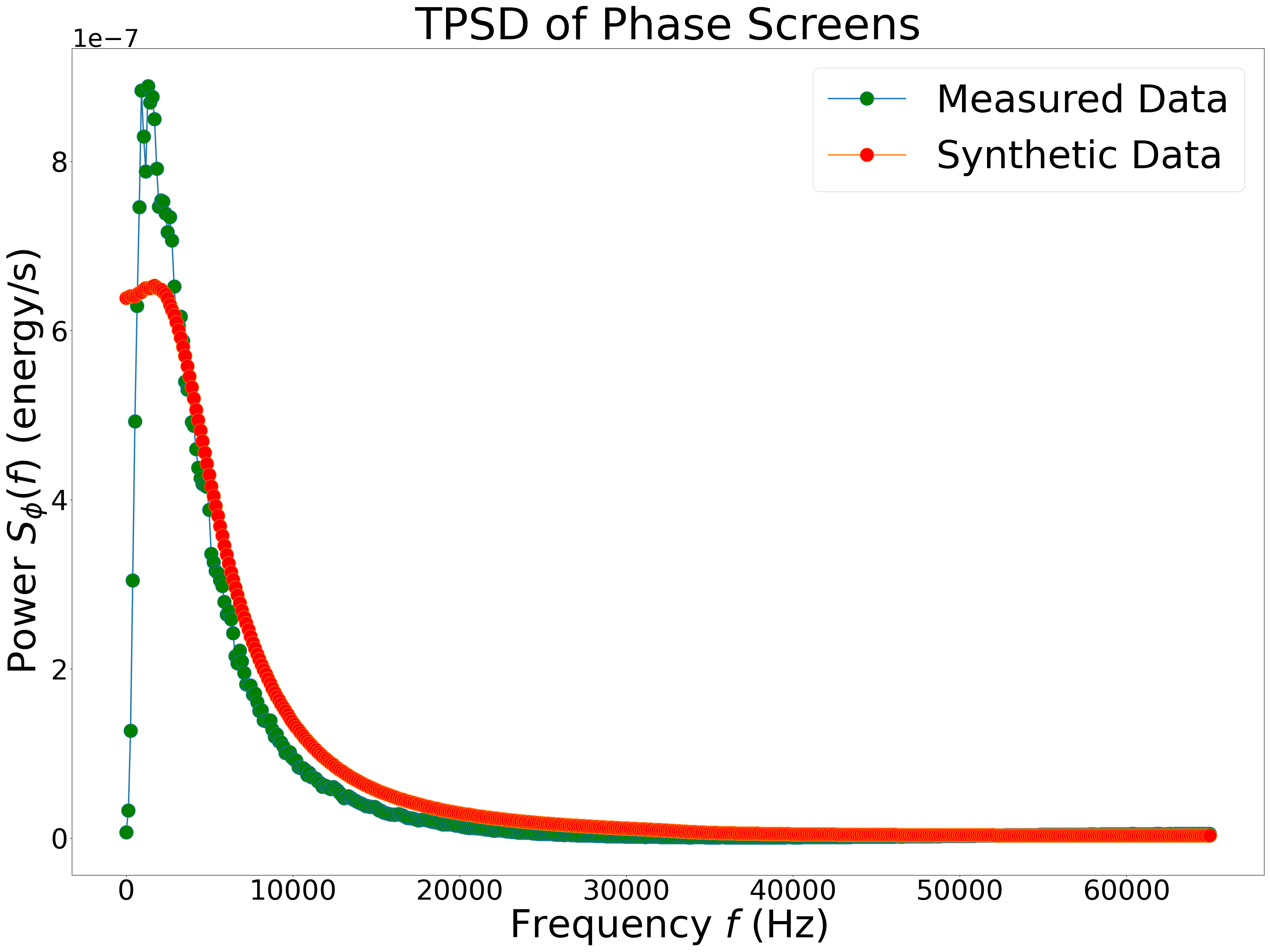}
        \caption{Data Set F12: TPSD of $\phi$}
        \label{fig: F12 TPSD of Phase}
    \end{subfigure}
    \caption{This figure shows temporal power spectral density (TPSD) plots of the measured data (blue) and synthetic data (orange). We show the TPSD of both the deflection angle in streamwise direction of the flow, $\theta_x$ (left), and the phase aberrations $\phi$ (right). Figures~\ref{fig: F06 TPSD of Deflection Angle} and~\ref{fig: F06 TPSD of Phase} show the TPSD of $\theta_x$ and $\phi$ for data set F06, while Figs.~\ref{fig: F12 TPSD of Deflection Angle} and~\ref{fig: F12 TPSD of Phase} show the TPSD of $\theta_x$ and $\phi$ for data set F12 (respectively). Temporal frequency $f$ (Hz) is plotted on the $x$-axis and the TPSD value $S(f)$ (energy/sec) is plotted on the $y$-axis. These plots illustrate that boiling flow reasonably matches the TPSD of $\theta_x$ at all frequencies and closely matches the TPSD of $\phi$ at frequencies above 20 kHz. However, it does match the TPSD of $\phi$ at frequencies below 20 kHz, nor does it match the lowest frequencies of either TPSD calculation. Thus, while our algorithm's direct estimation of the flow velocities $(v_x, v_y)$ fits the temporal statistics of the slopes of convective data, the three parameters $(v_x, v_y, \alpha)$ are unable to model the long-range temporal statistics of the measured phase aberrations.}
    \label{fig: TPSD}
\end{figure}

Figure~\ref{fig: TPSD} shows plots of the TPSD of $\theta_x$ and $\phi$ for both the measured data and synthetic phase screens. We see from Figs.~\ref{fig: F06 TPSD of Deflection Angle} and~\ref{fig: F12 TPSD of Deflection Angle} that boiling flow reasonably matches the TPSD of $\theta_x$, especially at the peak locations. This occurs because our algorithm directly fits the slopes of the phase aberrations $(v_x, v_y)$ from data; furthermore, since the data is highly convective, this method also fits the temporal statistics of the slopes. Furthermore, Figs.~\ref{fig: F06 TPSD of Phase} and~\ref{fig: F12 TPSD of Phase} show that the TPSD of the synthetic phase screens closely matches the TPSD of the measured data at frequencies above 20 kHz. However, boiling flow does not match the TPSD of the measured data at frequencies below 20 kHz. This likely occurs because boiling flow only uses three parameters $(\alpha, v_x, v_y)$ to fit the temporal statistics of the measured data and only uses a single time-lag in its prediction of each phase screen. These results suggest that these three parameters do not fully model the temporal statistics of the data; they are specifically unable to accurately model long-range temporal statistics. Furthermore, the TPSD of both $\theta_x$ and $\phi$ from the synthetic data have non-zero D.C. components because boiling flow has no mechanism to fit low-frequency temporal statistics.

Table~\ref{tab: Error Metrics} lists the error metrics for each data set. These results indicate that boiling flow reasonably matches the TPSD of $\theta_x$, with NRMSE below 9\% across both data sets. Furthermore, the NRMSE between the TPSD of the synthetic phase screens and the measured data is relatively low (at most 12.047\%). However, the NRMSE between the structure functions of the synthetic phase screens and the measured data illustrates that boiling flow does not accurately model the spatial statistics of aero-optic phase aberrations (with errors above 28\%). This suggests that boiling flow requires a more generalized model of the spatial statistics to generate aero-optically relevant phase screens.

\begin{table}[htbp]
    \caption{Data Set F06: Error Metrics}
    \label{tab: Error Metrics}
    \centering
    \begin{tabular}{|c||c|c|}
        \hline
        Error Metric & Data Set F06 & Data Set F12 \\
        \hline
        $NRMSE(S_{\theta_x},\hat{S}_{\theta_x})$ & 0.08919 & 0.08745 \\
        $NRMSE(S_{\phi}, \hat{S}_{\phi})$ & 0.09708 & 0.12047 \\
        $NRMSE(\bm{D}^{1/2}_{\phi/\sigma}, \bm{\hat{D}}^{1/2}_{\phi/\sigma})$ & 0.32584 & 0.28606 \\
        \hline
    \end{tabular}
\end{table}

\section{CONCLUSIONS}
\label{s: Conclusions}
In this paper, we introduced an algorithm for estimating the parameters of a well-known atmospheric phase screen generation method, boiling flow, from measured aero-optic phase aberration data. This algorithm estimates the outer scale $L_0$, Fried coherence length $r_0$, flow velocities $(v_x, v_y)$, and boiling parameter $\alpha$ to fit the spatial and temporal statistics of the measured data. Specifically, we estimated $(L_0, r_0)$ to fit the spatial statistics and $(v_x, v_y, \alpha)$ to fit the temporal statistics. 

We tested this method by estimating these parameters from two highly convective turbulent boundary layer (TBL) data sets containing measured aero-optic phase aberrations \cite{Kemnetz} and then using the parameters to generate phase screens. We evaluated the accuracy of this method by comparing the temporal power spectral density (TPSD) and Kolmogorov spatial structure function of the measured data with those of the (synthetic) phase screens. Our results show that the TPSD of the slopes of synthetic phase screens (i.e., the deflection angle $\theta_x$, taken in the streamwise direction of the flow) reasonably match that of the measured phase aberrations (see Figs.~\ref{fig: F06 TPSD of Deflection Angle} and~\ref{fig: F12 TPSD of Deflection Angle}), with errors between 8-9\% for both data sets. This likely occurs since the algorithm directly fits the slopes of the measured data, $(v_x, v_y)$. Further, this result suggests that our parameter estimation algorithm can fit the temporal statistics of the slopes of convective aero-optic phase aberration data. However, since this phase screen generation algorithm uses the Taylor frozen-flow method, it likely requires that the phase aberration data is highly convective to accurately fit the TPSD of the slopes of the data. Moreover, the TPSD of the synthetic phase screens themselves match that of measured phases screens slightly less closely (see Figs.~\ref{fig: F06 TPSD of Phase} and~\ref{fig: F12 TPSD of Phase}), with errors reaching 12\% for one data set. Because boiling flow only uses three parameters to fit the temporal statistics, $(v_x, v_y, \alpha)$, it likely does not fully capture the measured data's temporal statistics. Specifically, given the inaccurate TPSD fit at low-frequencies, these parameters do not accurately model the long-range temporal statistics of the measured data. 

Finally, we generalized the Kolmogorov spatial structure function, which we call the anisotropic structure function, to compute the mean-squared phase difference as a function of both the separation distance $r$ and angle $\theta$ between two sub-aperture locations. This generalization allowed us to evaluate the structure function of anisotropic data. Figures~\ref{fig: F06 Structure Function} and~\ref{fig: F12 Structure Function} show the anisotropic structure function of the synthetic phase screens and the measured phase aberrations, revealing that they do not match; in fact, the errors were above 28\% for both TBL data sets. Thus, the boiling flow parameters $(r_0, L_0)$ do not accurately fit the spatial statistics of aero-optic phase aberrations. This validates the assertion that the Kolmogorov theory of turbulence is inaccurate for aero-optic phase aberrations \cite{Vogel, Siegenthaler} and suggests that a more general model of the spatial statistics of phase screens is necessary for boiling flow to accurately simulate aero-optic phase aberrations. 

Future steps in this work include (1) defining a more accurate model of the spatial statistics of aero-optic phase aberrations and (2) testing this approach on data from AAOL experiments \cite{JumperAAOL}. The AAOL experiments measured additional aero-optic effects beyond a TBL. Importantly, the resulting phase aberration data sets are not necessarily convective, which poses an additional challenge for boiling flow.

\section*{DISCLAIMER}
The views expressed are those of the author and do not necessarily reflect the official policy or position of the Department of the Air Force, the Department of Defense, or the U.S. government. Approved for public release; distribution is unlimited.  Public Affairs release approval \# \PAnumber.  

\bibliography{report}
\bibliographystyle{spiebib}

\end{document}